\documentclass[journal=nalefd,manuscript=article]{achemso}
\usepackage[version=3]{mhchem}
\usepackage{xcolor,soul}
\usepackage{ulem}

\usepackage{upgreek}
\usepackage{xr-hyper}
\usepackage[hidelinks]{hyperref}
\usepackage{amssymb}

\author{Jitender Kumar}
\affiliation[WIS1]
{Department of Chemical and Biological Physics, Weizmann Institute of Science, Rehovot, Israel}
\author{Dan Yudilevich}
\affiliation[WIS1]
{Department of Chemical and Biological Physics, Weizmann Institute of Science, Rehovot, Israel}
\author{Ariel Smooha}
\affiliation[WIS1]
{Department of Chemical and Biological Physics, Weizmann Institute of Science, Rehovot, Israel}
\author{Inbar Zohar}
\affiliation[WIS1]
{Department of Chemical and Biological Physics, Weizmann Institute of Science, Rehovot, Israel}
\author{Arnab K. Pariari}
\affiliation[WIS2]
{Department of Condensed Matter Physics, Weizmann Institute of Science, Rehovot, Israel}
\author{Rainer St\"ohr}
\affiliation[USTUTT]{3.\,Physikalisches Institut, Universit\"at Stuttgart, Stuttgart, Germany}
\author{Andrej Denisenko}
\affiliation[USTUTT]{3.\,Physikalisches Institut, Universit\"at Stuttgart, Stuttgart, Germany}
\author{Markus H\"ucker}
\affiliation[WIS2]
{Department of Condensed Matter Physics, Weizmann Institute of Science, Rehovot, Israel}
\author{Amit Finkler}
\email{amit.finkler@weizmann.ac.il}
\affiliation[WIS1]
{Department of Chemical and Biological Physics, Weizmann Institute of Science, Rehovot, Israel}

\title{Room temperature relaxometry of single nitrogen-vacancy centers in proximity to  $\alpha$-RuCl$_3$ nanoflakes}


\begin{document}

\begin{tocentry}
    \includegraphics[width = 1.0\textwidth]{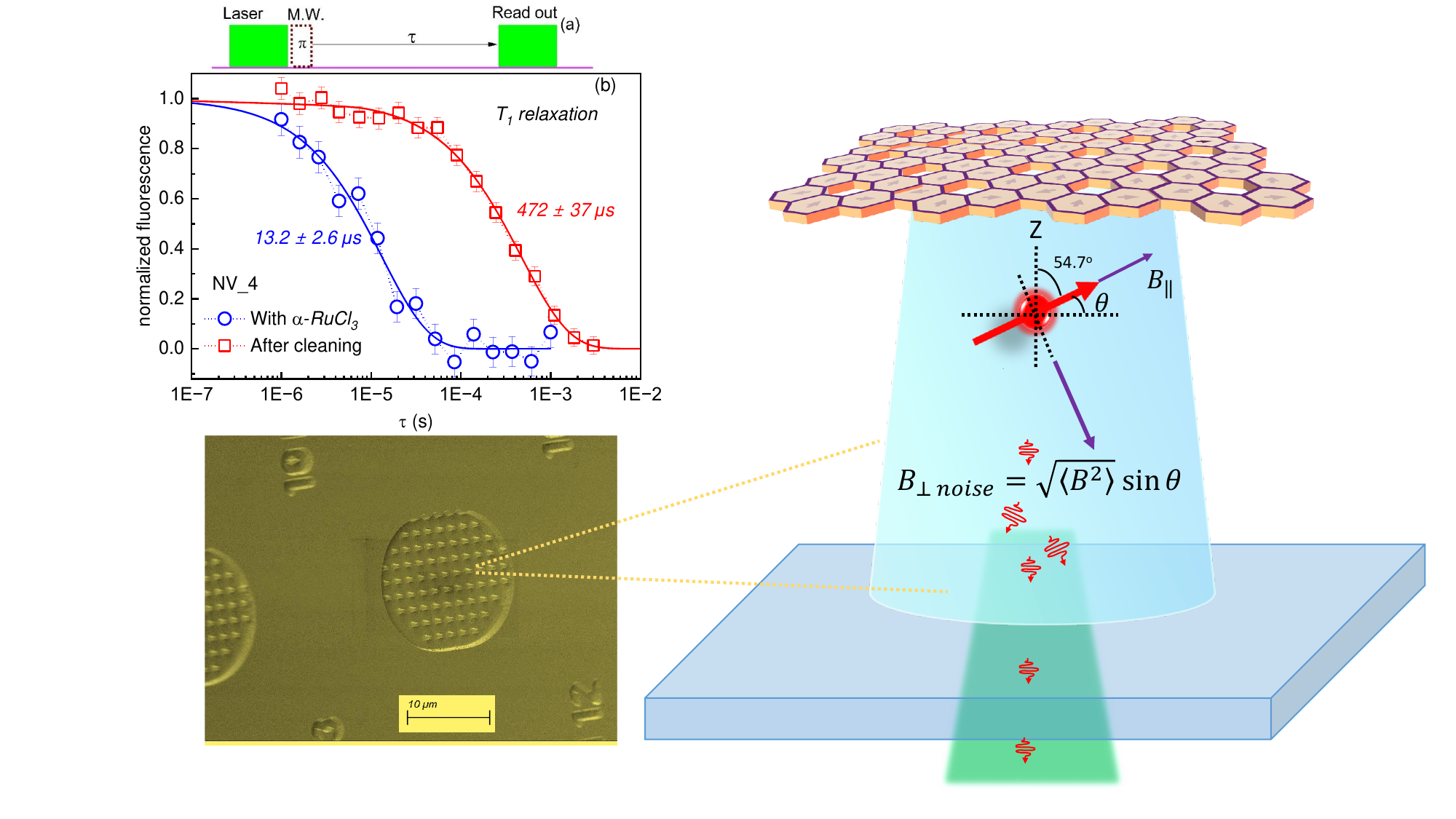}
\end{tocentry}

\begin{abstract}

Investigating spin and charge noise in strongly correlated electron systems is a valuable way to analyze their physical properties and unlock new phases of matter. In this context, nitrogen-vacancy (NV) center-based magnetometry has been proven to be a versatile sensor for various classes of magnetic materials in broad temperature and frequency ranges. Here, we use longitudinal relaxation time $T_1$ of single NV centers to investigate the spin dynamics of nanometers-thin flakes of  $\alpha$-RuCl$_3$ at room temperature. We observe a  significant reduction in the $T_1$ in the presence of $\alpha$-RuCl$_3$ in proximity to our NVs, which we attribute to paramagnetic spin noise confined in the 2D hexagonal plane. Furthermore, the $T_1$ time exhibits an almost linear increase with an applied external magnetic field. We associate this trend with the alteration of spin and charge noise in $\alpha$-RuCl$_3$ under an external magnetic field. These findings suggest that the influence of the room-temperature spin dynamics of $\alpha$-RuCl$_3$ on the longitudinal relaxation time of the NV center can be used to gain information on the material itself and the technique to be used on other 2D materials.  
\end{abstract}

\section{Introduction}
The NV center is a defect in the diamond lattice that forms an electronic qubit with a sufficiently long coherence time for a variety of applications, ranging from quantum computing and quantum information to quantum sensing \cite{Neumann2010, single, Two_qubit, Balasubramanian2008, Degen}. In the realm of sensing, the NV center serves as a highly sensitive sensor for many physical quantities, including temperature, rotation, electric and magnetic fields, and pressure \cite{RevModPhys,condensedmatt}. In magnetic field sensing, the NV center is susceptible not only to AC and DC fields but also to magnetic fluctuations across a broad frequency spectrum \cite{Fluctuating}. For instance, the transverse coherence times $T_2^*$ and $T_2$ of the NV center are susceptible to DC to MHz magnetic noise, depending on the pulse sequences used, such as Ramsey, Hahn echo, and CPMG \cite{Hahn_cpmg}. Complementing this, noise generated from fluctuations in the GHz regime affects the longitudinal relaxation time or spin-lattice relaxation time $T_1$ of the NV center \cite{Wideband}.\

Magnetic noise sensing finds a useful application in the study of 2D materials. Indeed, intense research has been directed toward van der Waals (vdW) strongly correlated materials in condensed matter physics. A weak interlayer vdW coupling and strong anisotropic magnetic interactions allow 2D materials to host interesting quantum properties in exfoliated single to few atomic layers such as layer-dependent magnetism (ferro/anti), multiferroicity, superconductivity and quantum spin liquid state \cite{2D_mag, layer, magnetic_genome, Banerjee}. Conventional experimental techniques widely used to explore low-dimensional magnetic systems include spin-polarized scanning tunneling microscopy \cite{Chen2019}, scanning SQUID magnetometry \cite{Zhou2023}, X-ray magnetic circular dichroism \cite{BedoyaPinto2021}, and neutron scattering \cite{Zhu2021}. Common to these techniques are the requirement of low temperatures and limited detection bandwidth for magnetic fluctuations. Moreover, in some, for instance the routinely used magnetic probe neutron scattering technique, a relatively large amount of sample is required, and it is also susceptible to high absorption in heavy elements, such as iridium. In recent studies on 2D magnetic materials, single NV center $T_1$ relaxometry has emerged as a vital noninvasive experimental tool for exploration of diverse physical properties at the nanoscale \cite{Science,condensedmatt,spinwave,collinear,Romana,Proposal,Demler}.\\

Recently, the 2D magnetic insulator, ruthenium trichloride ($\alpha$-RuCl$_3$), was proposed as a potential candidate to host an exotic state of matter: a quantum spin liquid (QSL) state. In QSLs, strong quantum fluctuations inhibit long-range spin order down to very low temperatures and instead form an unconventional ground state exhibiting long-range spin entanglement and topological order \cite{QSL, Naturereview, Review, Banerjee}. In 2006, Kitaev proposed an exactly solvable model for the QSL ground state in a honeycomb lattice, which supports the existence of fractional excitations like Majorana fermions and $\mathbb{Z}_2$ gauge fluxes \cite{Kitaev_1}. Besides the exciting physics of QSLs, they also have application potentials. For instance, long-range entangled spins have the potential to establish a quantum communication network, and Majorana fermions may serve as error-protected qubits in a future quantum computing technology \cite{Kitaev_2, com}.\newline
While $\alpha$-RuCl$_3$ displays intriguing magnetic phases at low temperatures, its high-temperature spin dynamics, particularly under external perturbation, is equally fascinating. For instance, the collapse of the Mott state and magnetic susceptibility has been observed at room temperature under external pressure\cite{collapse_1,collapse_2}. In addition, a scattering continuum, argued to be associated with fractional excitations has been observed in room temperature polarization-resolved Raman spectra of $\alpha$-RuCl$_3$ single crystals \cite{RT_Raman}. Considering the NV center’s excellent sensitivity to magnetic and electric field noise and drawing inspiration from recent theoretical proposals to employ $T_1$ relaxometry for exploring different phases of magnetic systems \cite{P_lee, Inti, Demler}, there is significant interest in utilizing NV relaxometry to explore nanoflakes of $\alpha$-RuCl$_3$.

In this work, we performed room temperature $T_1$ relaxometry on single NV centers to investigate the magnetic noise originating from nanometers-thin layers of $\alpha$-RuCl$_3$. Our results reveal a significant reduction in $T_1$ in the presence of these flakes. Additionally, we examined $T_1$ under varying external magnetic fields. Our findings illustrate that external magnetic fields alter the noise profile of $\alpha$-RuCl$_3$ and appear to enhance the $T_1$ of the proximate NV center.\\
\begin{figure}[ht!]
  \centering
  \includegraphics[width=1.0\textwidth]{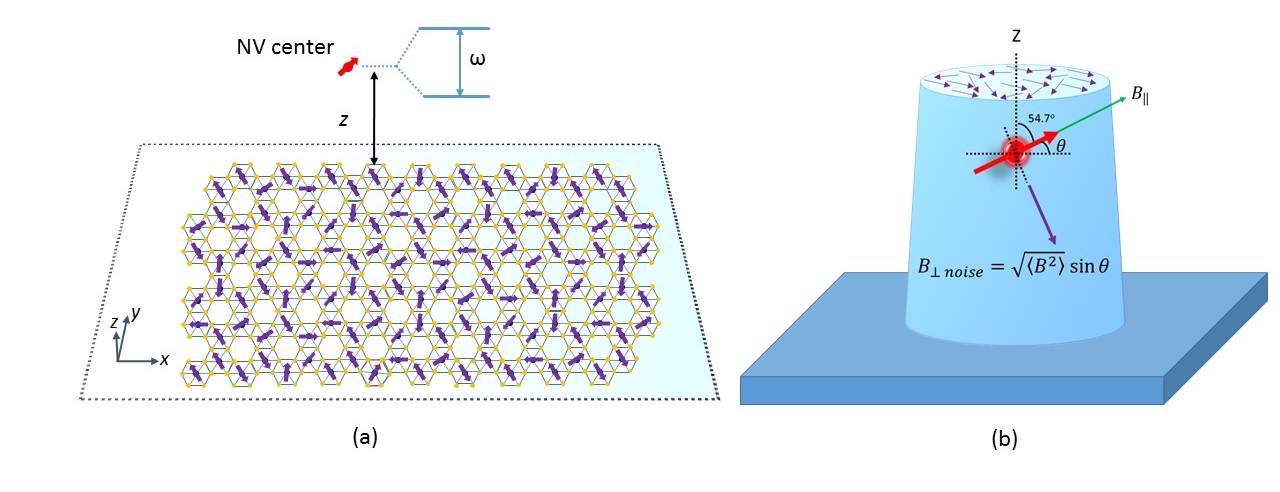}
  \caption{(a) Schematic of our experiment  depicting, a quantum sensor located at distance $z$  from a 2D $\alpha$-RuCl$_3$ spin bath. Ruthenium trichloride acquires a monoclinic structure (space group C2/m) at room temperature, where Ru$^{3+}$ ions form a 2D hexagonal honeycomb lattice in the \textit{ab} plane \cite{mono}. (b) Depiction of the projection of 2D noise, with the external magnetic field $B_\parallel$ aligned parallel to the NV quantization axis, for an NV center located in the diamond pillar.}
  \label{fig:1}
\end{figure}
\section{Results and discussion}

\begin{figure}[ht!]
  \centering
  \includegraphics[width=0.5\textwidth]{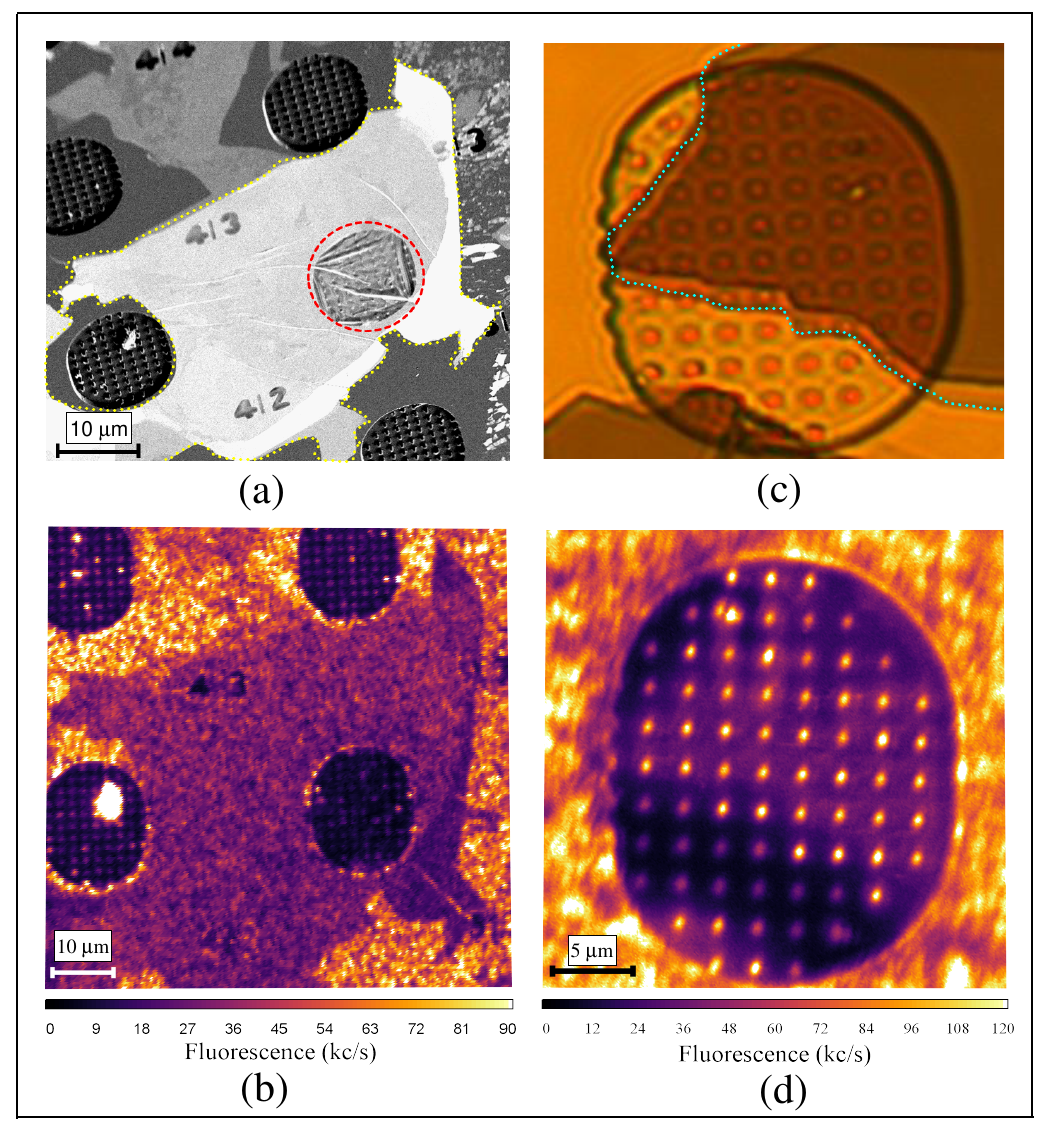}
  \caption{(a) 45-degree tilted SEM image of one of the \textbf{thin} $\alpha$-RuCl$_3$ flakes transferred on diamond outlined by dotted yellow line; the flake forms a tent-like structure over the nanopillars (encircled by a red dashed line). (b) A confocal microscope image of the same flake which shown in (a). The region covered with the flake exhibits less fluorescence, indicating quenching of the fluorescence in the presence of the atomically thin flake. (c) Optical microscope image of a $\sim$ 100\,nm thick flake  on the pillars, outlined by dotted line. (d) Confocal  image of the \textbf{ $\sim$ 100\,nm thick} flake shown in (c). The pillars beneath the flake are brighter than the pillars outside of the flake, which may result from the interaction between the flake and the fluorescing defect.}
  
  \label{fig:2}
\end{figure}
All fluorescence experiments were performed using a custom-built confocal microscope. Illustrated in Fig.\,\ref{fig:1}a, a spin qubit is positioned at a distance $z$ from a 2D dimensional material, while Fig.\,\ref{fig:1}b depicts an optically accessible  shallow NV center qubit in a diamond nanopillar. We start out with diamond membranes with shallow single NV centers (see SI for details); by using e-beam lithography, nanopillars which increase the photon collection efficiency are etched in the diamond. Since placing flakes on top of nanopillars is challenging, we developed a method of a hybrid bulk-nanopillar diamond structure to explore atomically thin flakes using single NV centers in pillars. In this hybrid structure, we created oval-shaped regions of nanopillars covering nearly 25$\%$ of the flat diamond surface area. Hence, one can enjoy the benefits of both cases, i.e., the flat areas providing strong adhesion, helping the flakes to adhere to and encircle the nanopillars seamlessly, while the nanopillars can host a single NV center and offer high photon counts per second. An SEM image of bare diamond pillars is provided in the SI (Fig.\,\ref{Fig.S1}).\

To investigate $\alpha$-RuCl$_3$ using single NV center $T_1$ relaxometry, we exfoliated thin flakes from a sufficiently large single crystal (see SI for details on crystal growth). These flakes were then transferred onto a $(100)$ diamond membrane patterned with nanopillars hosting NV centers using the Scotch tape dry exfoliation method. In Fig.\,\ref{fig:2}a, the SEM image illustrates that the transferred flake adopts a tent-like structure over the nanopillars. This pattern is similar to h-BN flakes transferred onto diamond nanopillars for artificial manipulation of curvature and strain \cite{Tent}. The thickness of the transferred flakes, as determined by atomic force microscopy (AFM), ranges from approximately 10\,nm to several hundred nm. The AFM height profile of the thinnest flake ($\sim$ 11\,nm) examined in this study is provided in Fig.\,\ref{Fig.S3}c of the SI. Optical microscope images reveal the thickness-dependent contrast of the transferred flakes. A relatively thick flake, approximately 100\,nm in thickness, is easily distinguishable atop the pillars in optical microscope images, as illustrated in Fig.\,\ref{fig:2}c.

In our experimental setup, we observed  up to approximately five times enhancement in the collection efficiency for nanopillars compared to flat diamond\cite{Maletinsky2012, Momenzadeh2015, Zhu}. High photon collection efficiency improves sensitivity and shortens measurement times for sequences incorporating $T_1$ \cite{Degen}. Another advantage of using pillars is that they allow us to pass the laser through the bottom of the diamond to illuminate the NV centers, which minimizes the laser impact on the material on top. This becomes particularly important in the case of atomically thin materials, where the higher laser power can damage the material \cite{Etching}. Thus, the utilization of nanopillars reduces the total irradiated energy on the sample due to the combined effects of faster measurements, reduced transmission to the sample, and lower laser powers.\\
\begin{figure}[ht!]
  \centering
  \includegraphics[width=.5\textwidth]{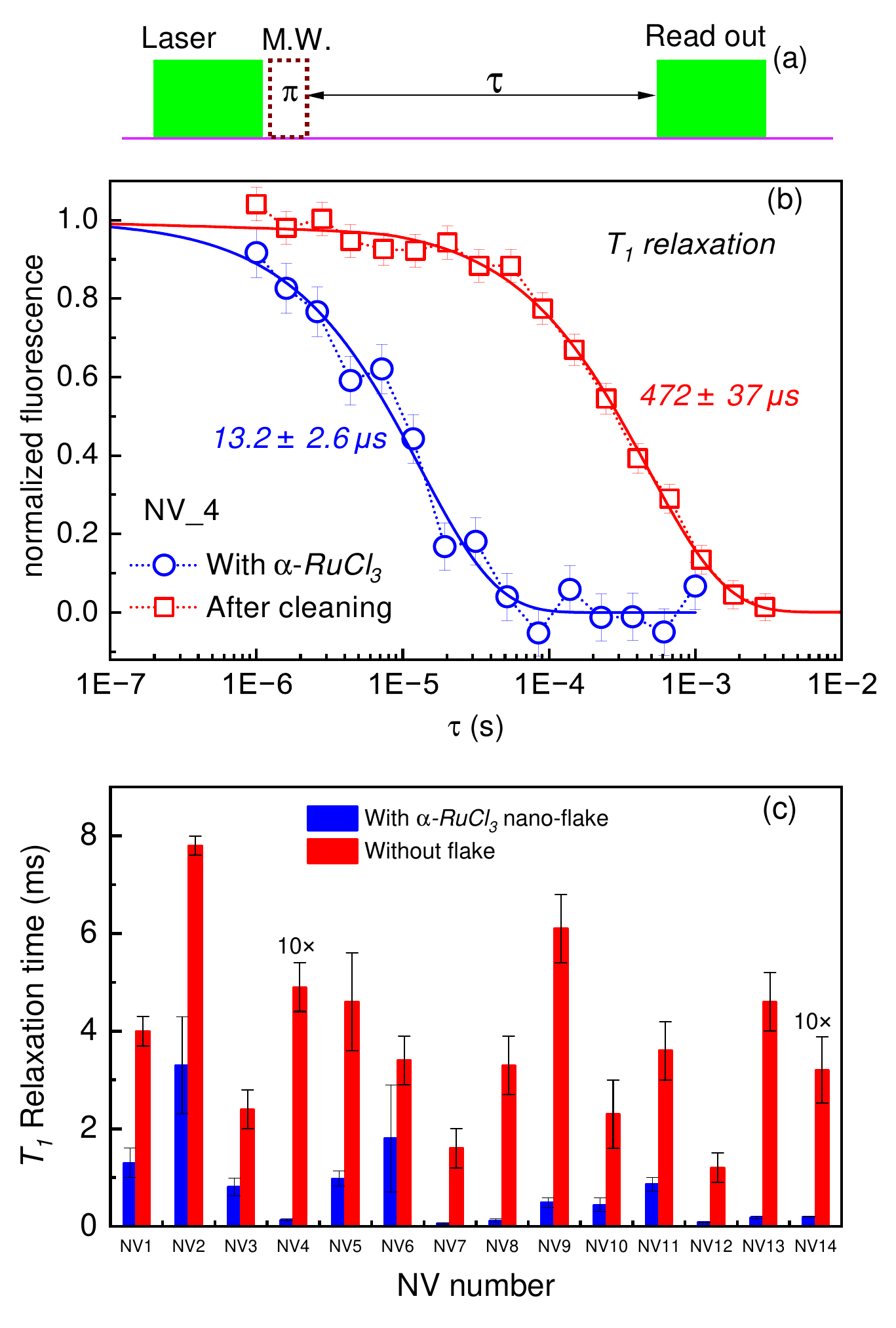}
  \caption{(a) Basic pulse sequence for a  typical $T_1$ measurement: The spin state $m_s=0$ is polarized with a laser pulse. The state interacts with noise bath during the waiting time and read out by the second laser pulse (b) Comparison of the  $T_1$ relaxation time decay of a single NV center measured in the presence (blue circles) and absence (red squares) of an 11 nm thick flake of $\alpha$-RuCl$_3$. Solid lines are stretched exponential fits explained in the text. (c) $T_1$ relaxation time measured for a set of single NV centers in the presence (blue) and absence (red) of thick (100 nm, NV12 to NV14) and thin flake ($\sim$ 11 nm, NV1 to NV11) of $\alpha$-RuCl$_3$. The values of NV4 and NV14 are multiplied by a factor of 10 to make them visible on the $y$ scale. }
  \label{fig:3}
 \end{figure}\\
To explore magnetic noise from $\alpha$-RuCl$_3$ flakes, we performed $T_1$ relaxometry of thin flakes of $\alpha$-RuCl$_3$ using a single NV center. The NV is polarized in the spin state $m_s=0$ using a 520\,nm laser, and this state freely decays to a thermal admixture of $m_s=0,\pm 1$. 
The decay rate of the polarized state depends on the coupling of the NV center to its environment. In general, the relaxation time of the NV center in the presence of spin noise has been described as\cite{Wideband}\ 
\begin{align*}
\frac{1}{T_{i}}=\left ( \frac{1}{T_{i}} \right )_\mathrm{int}+\int \gamma ^{2}\left \langle B \right \rangle^{2}S\left ( \omega ,T,E_{a} \right )F_{i}\left ( \omega  \right )d\omega,
\end{align*}
 where $\left(T_i^{-1}\right)_\mathrm{int}$ refers to intrinsic relaxation mechanisms, $\gamma$ is the gyromagnetic ratio of the NV center, and ${B}$ is the root mean square value of the effective magnetic field at the NV position produced by the proximate spin bath configuration. $E_a$ is the energy of magnetic anisotropy.  $S(\omega)$ is the spectral density of the spin noise environment, and $F(\omega)$ is a filter function, which depends on the pulse sequence used (the filter function for $T_1$ is provided in the SI). The relaxation rate of an NV center [$\Gamma$=1/$T_1$], proximate to a conductive or insulating magnetic environment, is governed by spin noise associated with the imaginary part of the magnetic susceptibility ($\chi^{''}$) — in other words, the imaginary part of the magnetic field auto-correlation function, \cite{condensedmatt, Science, Khoo, P_lee, Inti}.
\begin{align*}
\frac{1}{T_{1}}=\frac{\mu_{B}^{2} }{2\hbar}\coth\left (\frac{\beta\hbar\omega  }{2}\right )\mathrm{Im}\chi _{B_{Z}B_{Z}}\left ( z,\omega  \right ),
\end{align*}
 where $z$ is the distance of the noise source to the NV center, and $\omega$ is the transition frequency of the NV center (the concept is schematically depicted in Fig.\,\ref{fig:1}a). The amplitude of the imaginary part of the magnetic susceptibility may be modified by different factors, e.g., orbital diamagnetic susceptibilities of spinon, chargon and the diamagnetic susceptibility of the emergent gauge field \cite{Khoo, P_lee, Inti}. The $T_1$ relaxation time of the NV center is susceptible to magnetic noise in the GHz regime ($\sim$3\,GHz) \cite{Wideband}, which is suitable for detecting the magnetic field's autocorrelation function (magnetic noise) generated by spin liquids \cite{Inti}, metals \cite{Jayich} and magnetic insulators\cite{Demler, Science}.\
 
 In Fig.\,\ref{fig:3}b, we compare collected room temperature photoluminescence signals of a single NV center against the variation in waiting time $\tau$ in the presence and absence of  $\sim$ 11 nm thick flake of $\alpha$-RuCl$_3$.  The signals show an exponential decay trend, indicating a relaxation of the NV center polarized state $m_s=0$ to mixed $m_s=0,\pm 1$ states. The characteristic decay time $T_1$ is calculated by fitting the signal with a stretched exponential function \cite{stretched}. 
\begin{align*}
{S}=\tau _{0}\exp\left ( -\frac{\tau }{T_{1}} \right )^\eta,
\end{align*}
where $S$ is the signal, $\tau_0$ is the prefactor, $\tau$ waiting time, $\eta$ refers to the stretching parameter, and $T_1$ is the longitudinal relaxation time. For NV4, the fits yield $T_1$$^\mathrm{Ru}$=13.2$\pm{2.6}$ $\upmu$s and $T_1$= 472$\pm{37}$ $\upmu$s with and without flake, respectively (the fitting procedure is given in the SI). The ratio of $T_1$/$T_1$$^\mathrm{Ru}$ $\sim$ 35; such a significant reduction of $T_1$ in the presence of a flake indicates a strong coupling of the NV spin dynamics to magnetic noise of $\alpha$-RuCl$_3$. Similar reductions in $T_1$ have been observed in the case of proximate magnetic noise sources; such as superparamagnetic Fe$_3$O$_4$ nanoparticles \cite{Fe3O4}, paramagnetic Gd$^{3+}$ spins \cite{Gd}, and ferritin nanomagnets \cite{ ferritin, Wideband}. At room temperature, $\alpha$-RuCl$_3$ is in a paramagnetic state, confirmed by fitting the Curie-Weiss law to the inverse dc magnetic susceptibility  (Fig.\,\ref{Fig.S5} of the SI). The notable reduction in $T_1$ in the presence of $\alpha$-RuCl$_3$ suggests that the paramagnetic noise bath of $\alpha$-RuCl$_3$ couples to the NV center. The incoherent dipole-dipole interaction between the NV center and the nearby thermally fluctuating spin bath accelerates the relaxation rate of the NV center. Furthermore, in polarized Raman spectroscopy of $\alpha$-RuCl$_3$, a broad scattering continuum with asymmetric Fano-line shapes is recorded, arising from two-dimensional magnetic scattering at room temperature \cite{Raman, RT_Raman}. In general, asymmetric Fano-line shapes typically originate from a coupling between discrete phonon modes and electronic states, signifying metallicity in phase-separated systems \cite{Raman, Fano_1, Fano_2}. Therefore, we argue that there is noise, similar to that in metals, being present in $\alpha$-RuCl$_3$, which can account for the reduction of $T_1$ of a single NV center.
 \begin{figure}[ht!]
  \centering
  \includegraphics[width=1.0\textwidth]{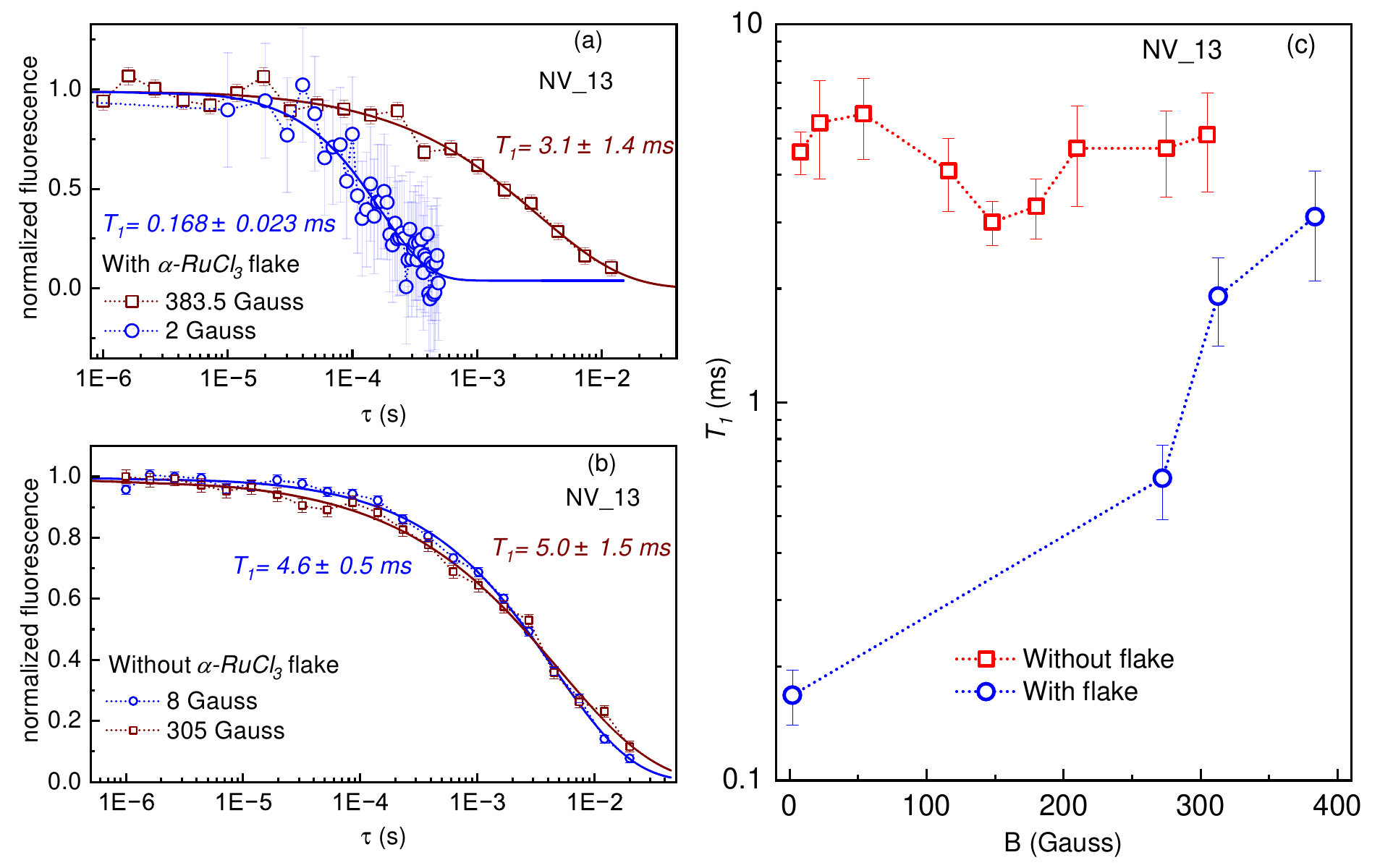}
  \caption{$T_1$ Magnetic field effect: Relaxation time of single NV center measured in the presence (a) and absence (b) of  $\alpha$-RuCl$_3$ flake for two different fields aligned parallel to the NV quantization axes. (c) $T_1$ relaxation versus magnetic field measured in the presence (blue) and absence (red) of the $\alpha$-RuCl$_3$ flake.}
  \label{fig:4}
\end{figure}\\
We have also collected $T_1$ data for NV centers covered by a relatively thick flake (100 nm), incorporated in Fig.\,\ref{fig:3}c as NV12, NV13, and NV14. The results closely resemble those observed under a thin flake. This behavior suggests that varying the thickness does not significantly alter the noise profile of the NV center. This is likely because the NV center is a highly localized probe, sensitive to its environment within  15 to 20\,nm radius. To validate this observation, we conducted $T_1$ experiments on a total of 14 individual NV centers, both with and without the presence of flakes. The results are shown in Fig.\,\ref{fig:3}c, with NV1 to NV11 representing thin flake conditions and NV12 to NV14 under thick flake conditions. The results consistently demonstrate a reduction of $T_1$  for each NV center in the presence of a flake. Investigating magnetic noise in the MHz regime, we utilized spin-echo spectroscopy ($T_2$ measurement), and the results are illustrated in Fig.\,\ref{Fig.S4} of the SI. The experiment was conducted on the same NV center as our $T_1$ relaxometry measurements in Fig.\,\ref{fig:3}b. Intriguingly, the transverse coherence time $T_2$ is barely affected by the presence of the flake, probably due to the low noise amplitude within this specific frequency range.\

To gain more insight into the magnetic behavior of the material at room temperature, we performed $T_1$ relaxometry under the application of an external magnetic field along the quantization axis of the NV center. This method is sometimes referred to as cross-relaxometry \cite{Temp_T1}. In a cross-relaxometry experiment, applying a magnetic field leads to Zeeman splitting, resulting in a linear opening of the energy gap between NV states $m_s=-1$ and $m_s=+1$. Each state exhibits distinct decay rates determined by the comparability of its energy with the transition rate of the nearby noise bath.\

Figure \ref{fig:4}a displays the $T_1$ of a single NV center, for two values of applied fields (2 and 383.5 Gauss) in the presence of flake. $T_1$ shows a stretched exponential decay and notably increases approximately 18-fold at the higher field. In contrast, the change was negligible when we collected data for different fields without the flake (Fig.\,\ref{fig:4}b). In Fig.\,\ref{fig:4}c, we summarize our field-dependent data for $T_1$. In the absence of the flake, the change in $T_1$ with respect to field variation is almost negligible, or the change is within the error. Remarkably, in the presence of the flake, $T_1$ exhibits an almost linear increase as the magnetic field strength increases, and at high field values, it nearly returns to the value observed in the absence of the flake. 
Such a significant increment of $T_1$ with magnetic field at room temperature is unusual due to the incompatible energy scales, i.e., $k_BT=25\,\mathrm{meV}$ and exchange interaction ($J\sim 3\,\mathrm{meV}$) \cite{Ran2017}, and so its origin is not immediately apparent. We discuss two possible mechanisms.\\
\textbf{A. Diamagnetism}. It has been observed that a diamagnetic electrolyte near the NV center reduces both electric and magnetic noise, consequently improving $T_1$ \cite{Fabian}. A theoretical calculation for the $T_1$ relaxation time of an NV center proximate to a 2D spin liquid shows similar effects of a diamagnetic state on the $T_1$ \cite{P_lee}.
It is worthwhile noting that a Curie-Weiss fit of the magnetic susceptibility gives a magnetic moment of 2.31 $\mu_B$. The inverse susceptibility data is fitted with a Curie-Weiss formula at a high-temperature regime to calculate the total paramagnetic moment (Fig.\,\ref{Fig.S5} in the SI). This value is significantly larger than the spin-only contribution of Ru$^{3+}$ in its low-spin state, which is 1.73 $\mu_B$\cite{orbital}, and indicates a significant contribution from the orbital moment. Interestingly, the calculated orbital moment is nearly one-third of the spin moment, resembling the free electron model, where the Landau diamagnetic response of the orbitals is precisely one-third of the Pauli paramagnetic response \cite{Landau}. In their recent theoretical work, Zin et al.\,\cite{diamag}. explored the intertwining of spin and orbital magnetization in $\alpha$-RuCl$_3$ using multiorbital spin-orbit model for Mott insulators such as $\alpha$-RuCl$_3$. They formulate that local current loops are induced when an external magnetic field is applied in such Mott insulators. This generates magnetization opposite to the applied field as a diamagnetic response similar to Landau diamagnetism in metals. Due to the substantial orbital moment present in $\alpha$-RuCl$_3$, we propose that applying an external magnetic field could induce a diamagnetic-like response in $\alpha$-RuCl$_3$, potentially leading to an enhancement in $T_1$.\\
\textbf{B. Charge redistribution.} In a recent work, a noticeable enhancement in the transverse coherence time of a NV centers has been observed in the presence of a superconductor \cite{SuperT2}. The effect is tentatively rationalized as a change in the electric noise due to the superconductor (diamagnetic)-induced redistribution of charge carriers near the NV center. $\alpha$-RuCl$_3$ is a small band-gap semiconductor with room temperature resistivity of the order of 10$^{-1}$ $\Omega$ cm\cite{Nano_raman}. In $\alpha$-RuCl$_3$, due to the onsite Coulomb repulsion, a gap opens within the Ru $4d$ band and splits it into two sub-bands, the upper Hubbard band and the lower Hubbard band, and an inter-site $d^5$$d^5$$\rightarrow$$d^4$$d^6$ charge transfer between neighboring Ru$^{3+}$ ions governs the hopping conduction mechanism \cite{Mott}. Along parallel lines as those of the change in transverse coherence time due to charge fluctuations, we therefore suggest that these could be an additional source of noise at room temperature. Charge fluctuations are sensitive to external perturbations, such as magnetic fields, especially in poor conductors. An external magnetic field may reduce the mobility of charges, leading to a decrease in charge fluctuations, and could enhance the $T_1$. The electrical conductivity of single crystal $\alpha$-RuCl$_3$ is highly anisotropic, with a reported conductivity almost four orders of magnitude higher in a plane perpendicular to the $c$ axis compared to the plane parallel to the $c$ axis\cite{aniso}. The longitudinal relaxation time of an NV center is sensitive to noise projected perpendicular to its quantization axis. In the presence of magnetic noise coupled to an NV center, the longitudinal relaxation rate can be expressed as \cite{art}.
\begin{align*}
\frac{1}{T_{1}}=\frac{1}{T_{0}}+3\gamma^{2}\sqrt{\left \langle B^{2} \right \rangle_{\perp NV }},
\end{align*}
where $\gamma$=28 GHz/Tesla is the electron gyromagnetic ratio, and $B$ is the magnetic field noise component perpendicular to the NV quantization axis.  The schematic of experimental geometry is shown in Fig.\,\ref{fig:1}b. The noise bath comprising of electronic spins and charge fluctuation lies in the ${ab}$ plane. The component of this noise perpendicular to the quantization axis of the NV center will be
\begin{align*}
\sqrt{\left \langle B^{2} \right \rangle_{\perp NV }}=\sqrt{\left \langle B^{2} \right \rangle} \sin\theta,
\end{align*}
where $\theta$ is the angle between the noise bath and NV quantization axis. Considering the two-dimensional nature of the spin-spin interactions in the hexagonal ${ab}$ plane, we believe that applying the magnetic field parallel to the NV quantization axis will diminish the angle $\theta$ and the noise component perpendicular to the NV quantization. This suggestion further validates our observation of $T_1$ recovery at a high field.

In conclusion, we have investigated nanometers-thick flakes of  $\alpha$-RuCl$_3$ using the longitudinal spin relaxation time $T_1$ of single NV centers in diamond. We observed a significant reduction of NV relaxation time by the proximity of $\alpha$-RuCl$_3$, which signifies the coupling of noise fluctuations of $\alpha$-RuCl$_3$ with the NV. The reduction in $T_1$ is almost independent of variations in the flake thickness, confirmed with 11-nm and 100-nm thick flakes. A gradual increment of $T_1$ time as a function of  magnetic field  in the flake's presence was observed. We hypothesize that the increase in $T_1$ is associated with the suppression of spin and charge fluctuations in atomically thin flakes due to the magnetic field. These results indicate that $T_1$ relaxometry with a single NV center is a promising technique to explore the spin dynamics of nanoflakes of potential quantum spin liquid candidate $\alpha$-RuCl$_3$ and similar systems
. A comprehensive low-temperature $T_1$  relaxometry study, along with supporting experiments such as magneto-resistance, are required to corroborate our interpretation. We also propose exploiting a (111) oriented scanning NV center to study the out-of-the-plane spin dynamics confined in the \textit{ab} hexagonal plane \cite{oriented}. Furthermore, unlike neutron scattering in  iridium containing compounds, the NV qubit does not suffer from issues related to flux absorption and amount of the sample, rendering it a desirable sensor for the family of iridium-based QSLs. 

\begin{acknowledgement}
We heartily appreciate comments from I.\,Sodemann on our manuscript. We thank N.\,Bar-Gill's group for allowing us to use their setup and H.\,Steinberg's group for showing us how to handle vdW materials. We also thank A.\,Cohen and O.\,Yaffe for running Raman characterization of our flakes.
MH is incumbent of the Henry and Gertrude F.\,Rothschild Research Fellow Chair and acknowledges support of the Leona M. and Harry B. Helmsley Charitable Trust grant no.\,2018PG-ISL006. AKP acknowledges postdoctoral fellowship support from the Council for Higher Education, Israel, 
through the Study in Israel program.
The authors acknowledge the historical support of the Perlman family as well as the Kimmel Institute for Magnetic Resonance Research. AF is the incumbent of the Elaine Blonde Career Development Chair in Perpetuity and acknowledges financial support from the Minerva Stiftung (grant no.\,714131) and the Israel Science Foundation (grant no.\,418/20).

\end{acknowledgement}

\bibliography{main}

\newpage

\end{document}